\documentclass[%
superscriptaddress,
preprint,
amsmath,
amssymb,
aps,
]{revtex4-2}

\usepackage{graphicx}
\usepackage{dcolumn}
\usepackage[export]{adjustbox}
\usepackage{bm}
\usepackage[caption=false]{subfig}
\usepackage{chemformula}
\usepackage{hyperref}
\usepackage{xcolor}
\usepackage{soul}

\begin{document}

\title{Metal Nanoparticle-Functionalized Three-Dimensional Graphene:\\a versatile platform towards sensors and energy-related applications}

\author{Emanuele Pompei}
\affiliation{NEST, Istituto Nanoscience-CNR and Scuola Normale Superiore, Piazza S. Silvestro 12, 56127, Pisa, Italy}

\author{Ylea Vlamidis}
\affiliation{NEST, Istituto Nanoscience-CNR and Scuola Normale Superiore, Piazza S. Silvestro 12, 56127, Pisa, Italy}
\affiliation{Department of Physical Science, Earth, and Environment, University of Siena, Via Roma 56, 53100, Siena, Italy}

\author{Letizia Ferbel}
\affiliation{NEST, Istituto Nanoscience-CNR and Scuola Normale Superiore, Piazza S. Silvestro 12, 56127, Pisa, Italy}

\author{Valentina Zannier}
\affiliation{NEST, Istituto Nanoscience-CNR and Scuola Normale Superiore, Piazza S. Silvestro 12, 56127, Pisa, Italy}

\author{Silvia Rubini}
\affiliation{Istituto Officina Dei Materiali IOM - CNR, Laboratorio TASC, Area Science Park, S.S.14, Trieste, I-34149, Italy}

\author{Daniel Arenas Esteban}
\affiliation{EMAT, University of Antwerp, Groenenborgerlaan 171, B-2020, Antwerp, Belgium}
\affiliation{Nanolab Centre of Excellence, Groenenborgerlaan 171, B-2020, Antwerp, Belgium}

\author{Sara Bals}
\affiliation{EMAT, University of Antwerp, Groenenborgerlaan 171, B-2020, Antwerp, Belgium}
\affiliation{Nanolab Centre of Excellence, Groenenborgerlaan 171, B-2020, Antwerp, Belgium}

\author{Carmela Marinelli}
\affiliation{Department of Physical Science, Earth, and Environment, University of Siena, Via Roma 56, 53100, Siena, Italy}

\author{Georg Pfusterschmied}
\affiliation{Institute of Sensor and Actuator Systems, TU Wien, 1040, Vienna, Austria}

\author{Markus Leitgeb}
\affiliation{Institute of Sensor and Actuator Systems, TU Wien, 1040, Vienna, Austria}

\author{Ulrich Schmid}
\affiliation{Institute of Sensor and Actuator Systems, TU Wien, 1040, Vienna, Austria}

\author{Stefan Heun}
\affiliation{NEST, Istituto Nanoscience-CNR and Scuola Normale Superiore, Piazza S. Silvestro 12, 56127, Pisa, Italy}

\author{Stefano Veronesi}
\affiliation{NEST, Istituto Nanoscience-CNR and Scuola Normale Superiore, Piazza S. Silvestro 12, 56127, Pisa, Italy}
\email{stefano.veronesi@nano.cnr.it}

\date{\today}

\begin{abstract}
We demonstrate the first successful functionalization of epitaxial three-dimensional graphene with metal nanoparticles. The functionalization is obtained by immersing the 3D graphene in a nanoparticle colloidal solution. This method is versatile and here is demonstrated for gold and palladium, but can be extended to other types and shapes of nanoparticles.
We have measured the nanoparticle density on the top-surface and in the porous layer volume by Scanning Electron Microscopy and Scanning Transmission Electron Microscopy. Samples exhibit a high coverage of nanoparticles with minimal clustering. High quality graphene has been demonstrated to promote the functionalization leading to higher nanoparticle density, both on the surface and in the pores. X-ray Photoelectron Spectroscopy allowed to verify the absence of contamination after the functionalization process. Moreover, it confirmed the thermal stability of the Au- and Pd-functionalized three-dimensional graphene up to 530°C. Our approach opens up new avenues for utilizing three-dimensional graphene as a versatile platform for catalytic applications, sensors, and energy storage and conversion.
\end{abstract} 

\maketitle

\section{\label{sec:level1}Introduction}

Inorganic porous materials have found a large variety of applications in the last decades. Their main feature is the high surface to volume ratio. Therefore, they are employed in the fields of catalysis \cite{Kadja2022}, fluids absorption \cite{Memetova2022}, sensors \cite{Singh2014}, energy storage and conversion \cite{Macili2023, Shao2020}, and drug delivery \cite{Mabrouk2019}.
Among all porous materials, those based on carbon have attracted a special interest. Carbon is a very versatile element since it is capable to bond to many different atoms, can rearrange in different crystalline structures, and possesses a high chemical stability and strength-to-density ratio. Porous C-based materials have been successfully produced with pore dimensions from less than 2~nm (micro-pores) \cite{Mohamed2021} to hundreds of nanometers (macro-pores) \cite{Funabashi2017}. Since the discovery of graphene in 2004 \cite{Novoselov2004}, effort has been put in finding a way to match the features of the previously realized porous C-based materials with the astonishing properties of graphene, such as high charge carrier mobility \cite{Bolotin2008}, large thermal conductivity \cite{Balandin2011}, high strength \cite{Lee2008}, and high specific surface area of 2630~m$^2$/g \cite{ZHANG2020127098}.

Many approaches have been developed to produce three-dimensional structures of graphene \cite{Ma2014}. Graphene foams (GFs) are made by chemical vapour deposition over a porous metallic template (typically made of nickel) \cite{Chen2011}. Graphene sponges (GSs) have a structure similar to GFs, but here the graphene flakes are partially oriented parallel to each other \cite{yu2015}. GSs are commonly made by freeze-drying of graphene oxide (GO) solutions \cite{Xu2015}. In addition, it is possible to realize graphene aerogels and hydrogels, in which GO flakes are induced to cross-link via chemical methods, resulting in three-dimensional graphene-like structures \cite{Worsley2010, Xu2010}. 

Recently, a new procedure for the production of three-dimensional graphene (3DG) has been demonstrated. It relies on the epitaxial growth of graphene on porous silicon carbide (SiC) \cite{Veronesi2022}. A 4H-SiC substrate is photoelectrochemically porousified \cite{Leitgeb2}, and the epitaxial growth of graphene is obtained via thermal decomposition of 4H-SiC. 3DG pore dimensions can be tuned from few nanometres to hundreds of nanometres by adjusting the SiC porousification process \cite{Leitgeb2017a}. It has been shown that graphene grows on the walls of the pores as a continuous film \cite{Veronesi2022}. This enhances the electrical conductivity of the material which can be fruitful in sensing applications \cite{veronesi2023}. 
The as-produced 3DG possesses many advantages compared to that obtained by other methods. Graphene quality is higher compared to GFs \cite{Chen2011}, and aerogels and hydrogels \cite{Gorgolis2017}. Owing to the SiC backbone, the 3DG is mechanically more stable compared to other structures. The doped bulk SiC below the porous layer can be used to electrically contact 3DG, which can be useful for implementations in the fields of sensors, supercapacitors, and fuel cells. In addition, its fabrication can be conducted on the wafer scale and is thus salable.

To further increase the versatility of this material, the functionalization with metals is an option. Most of the work done on metal-decorated graphene, both flat and three-dimensional, relies on electrodeposition \cite{Ma2014a, Peng2020}, direct chemical synthesis over the graphene support \cite{Toth2015, Nguyen2016}, and e-beam evaporation \cite{Mashoff2013}. The first two approaches do not guarantee a great control on the metal nanoparticle dimensions and distribution, and the process is usually not clean, i.e., contaminants present in the solution remain on the sample surface. The latter approach can be fine tuned in order to deposit a very precise amount of metal (even well below one monolayer), but it is unfeasible to deposit metal deep into a porous structure such as 3DG.
Here, we demonstrate a different approach for the functionalization of a structured graphene material. It consists in loading 3DG with metal nanoparticles (NPs) by immersing the 3DG into a colloidal solution with the NPs, similar to what has been done by Lepesant et al. for porous silicon \cite{Lepesant2018}.

\section{\label{results}Results}

\subsection{\label{sec:aunps}Functionalization with Gold Nanoparticles}

\begin{figure*}
\subfloat{\includegraphics[width=0.95\textwidth]{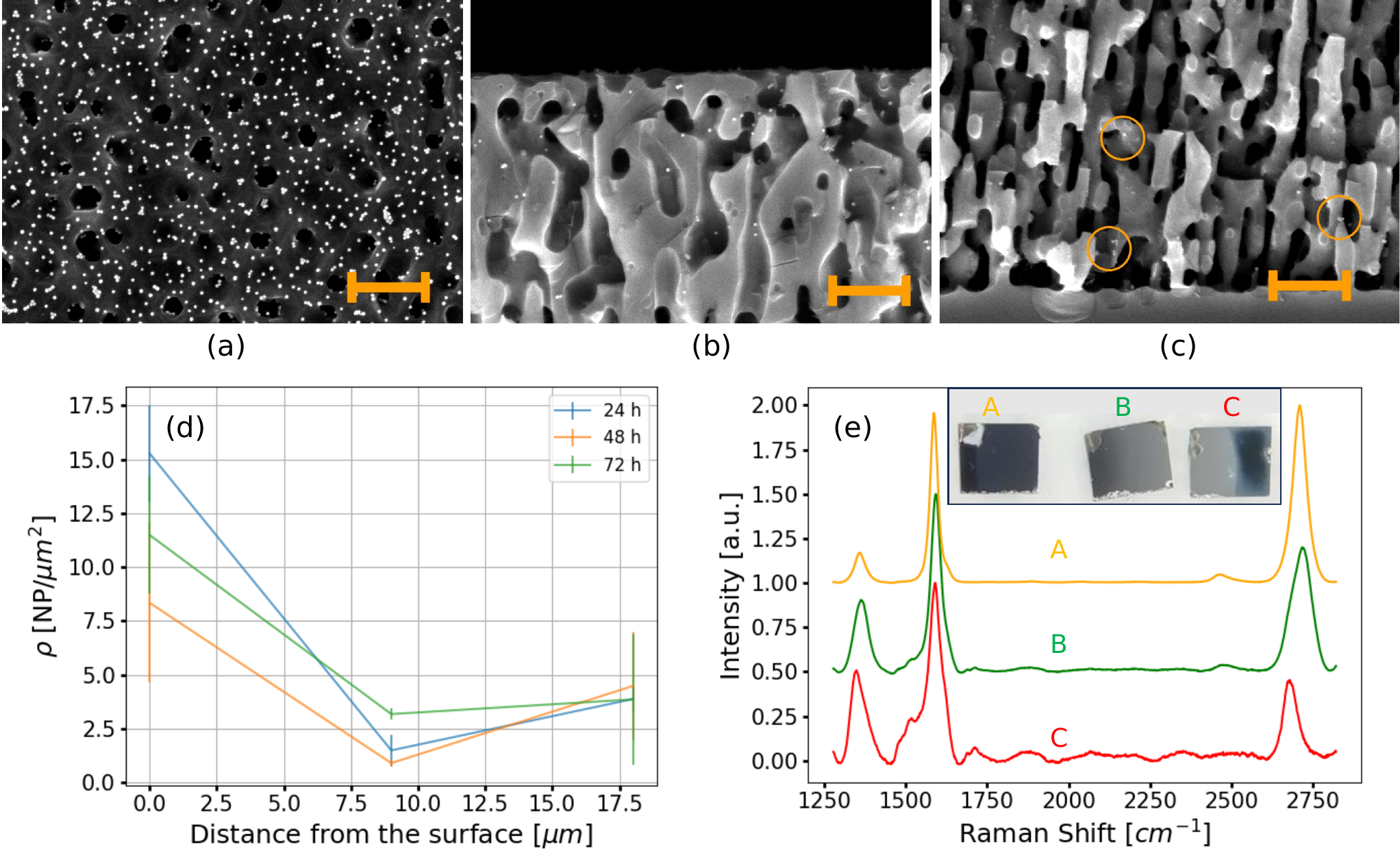}
}
\caption{\label{fig:a} (a) Top-view SEM image of an AuNP-functionalized 3DG sample. (b), (c) Cross-sectional SEM images of the top and the bottom region of the porous layer, respectively. In the image from the bottom region, some of the NPs are circled in orange. In all images, the orange scale bar indicates 0.5~$\mu$m. (d) Trend of the NP density in the pores at different depth, for different immersion times (24, 48, and 72 hours). (e) Raman spectra, acquired in the three distinct regions of the sample (A, B, and C), revealing a graphene quality gradient. The inset shows an optical image of the 3DG sample cleaved in three pieces, exhibiting a color gradient.}
\end{figure*}

As proof of concept for the functionalization of 3DG with metal nanoparticles, we selected commercially available, water suspended, spherical gold nanoparticles (AuNPs), with nominal diameter of 20 nm (purchased from BBI Solutions, Au content 0.01\% w/v). AuNPs not only are a common choice for plasmonic \cite{Amendola2017} and medical \cite{Giljohann2010} applications, but also guarantee sharp contrast over a carbonaceous substrate in SEM imaging (because of the large difference in atomic masses).

Since 3DG is highly hydrophobic, we transferred the nanoparticles from the initial water solution to ethanol. Then, a drop of the AuNPs colloidal solution was deposited over a 3DG sample. As the drop is released over the surface, it is immediately absorbed inside the pores. Therefore, we chose ethanol for the following functionalization experiments. 

The functionalization experiments were performed by immersing the 3DG samples in the NPs solution. The samples were then dried in air and analyzed by SEM, both on the top-surface and in cross-section. SEM measurements confirmed a successful functionalization, since NPs were detected both on the surface and inside the pores. The optimal results were achieved by immersing a 3DG sample in the NPs solution for 24 hours. Figure \ref{fig:a}a shows a SEM image of the surface of the gold-decorated 3DG sample. SEM measurements demonstrate that this approach was very effective, resulting in a nanoparticle top-surface density $\sigma$ of (220 $\pm$ 25)~NPs/$\mu$m$^2$ (obtained as the average over many SEM images). Nanoparticles were homogeneously distributed over the sample surface, and significant clustering was not observed (as visible from Fig.~\ref{fig:a}a). Figure \ref{fig:a}b and \ref{fig:a}c show typical SEM measurements acquired in the top and bottom regions of the porous layer. Interestingly, the density inside the pores, $\rho$, shows a dependence on the depth, i.e., the distance from the top-surface. In the first 500~nm below the surface, the density is $\rho_{top}=(15.2\pm1.2)$~NPs/$\mu$m$^2$. At half of the porosification depth, i.e., between about 8.75~$\mu$m and 9.25~$\mu$m, the density is $\rho_{mid}=(1.3\pm0.8)$~NPs/$\mu$m$^2$, and in the last 500~nm before the bulk SiC, the density is $\rho_{bot}=(3.7\pm1.0)$~NPs/$\mu$m$^2$. From these data we conclude that the NP density inside the pores decreases with depth, but there is also a tendency of NPs to accumulate at the bottom of the porous layer. This indicates that NPs could diffuse inside the pores for distance longer than the porousification depth.

Furthermore, if the 3DG sample is immersed in the NPs solution under ultrasonication, a much reduced NP density is obtained. This suggests that the vibration, instead of promoting the diffusion of NPs inside the pores, provides kinetic energy to the NPs sufficient to detach them from graphene.

Thus, we have established that a long immersion in the NPs solution in static condition is the best method for the functionalization of 3DG. Therefore, we tested longer immersion times of 48 and 72 hours. Figure \ref{fig:a}d shows the NP density inside the pores as a function of depth obtained for 24, 48, and 72 hours of immersion in the NPs solution. Since no significant increase in the nanoparticle density was observed, we concluded that 24 hours are sufficient for the NPs diffusion.

Graphenizing a sample in a non--homogeneous manner allowed us to study the role of the graphene quality on the functionalization. The non--homogeneous graphenization is achieved by inducing a thermal gradient along the sample in order to have one side at the optimal growth temperature and the opposite one more than 50 K colder (see Method Section). The sample has been functionalized by immersion in the NPs solution for 24 hours prior the cleavage. Then, it has been cleaved in three pieces labeled as "A", "B", and "C". The inset of Figure \ref{fig:a}e shows an optical picture of this sample (after cleavage), in which a color gradient is clearly visible. Besides, three Raman spectra, acquired one on each piece of the sample, are also shown. For each region, the shown spectrum is the average of all spectra acquired over a 21$\times$21 $\mu$m$^2$ map. Considering common Raman spectroscopy benchmarks of graphene such as the 2D vs. G peak intensity ratio and the G vs. D peak intensity ratio, it emerges that in "A" region, where the optimal temperature was achieved, high quality graphene is grown. The graphene quality degrades moving towards "C" (histogram of 2D vs. G peak intensity ratio in the different regions is shown in Fig.~S2). After the functionalization, we investigated by SEM the surface of "A" and "C" regions, as well as the cross-sections between "A" and "B", and "B" and "C". We observed a sharp drop in the NP densities from "A" (cf. \ref{fig:a}a--c) to "C": the top-surface density was about six times smaller in "C" ($\sigma\simeq34$~NPs/$\mu$m$^2$), and the density inside the pores was reduced by about a factor four to $\rho_{top}\simeq3.9$~NPs/$\mu$m$^2$. Thus, high graphene coverage and quality promote the functionalization process.

\begin{figure*}
\subfloat{\includegraphics[width=0.92\textwidth]{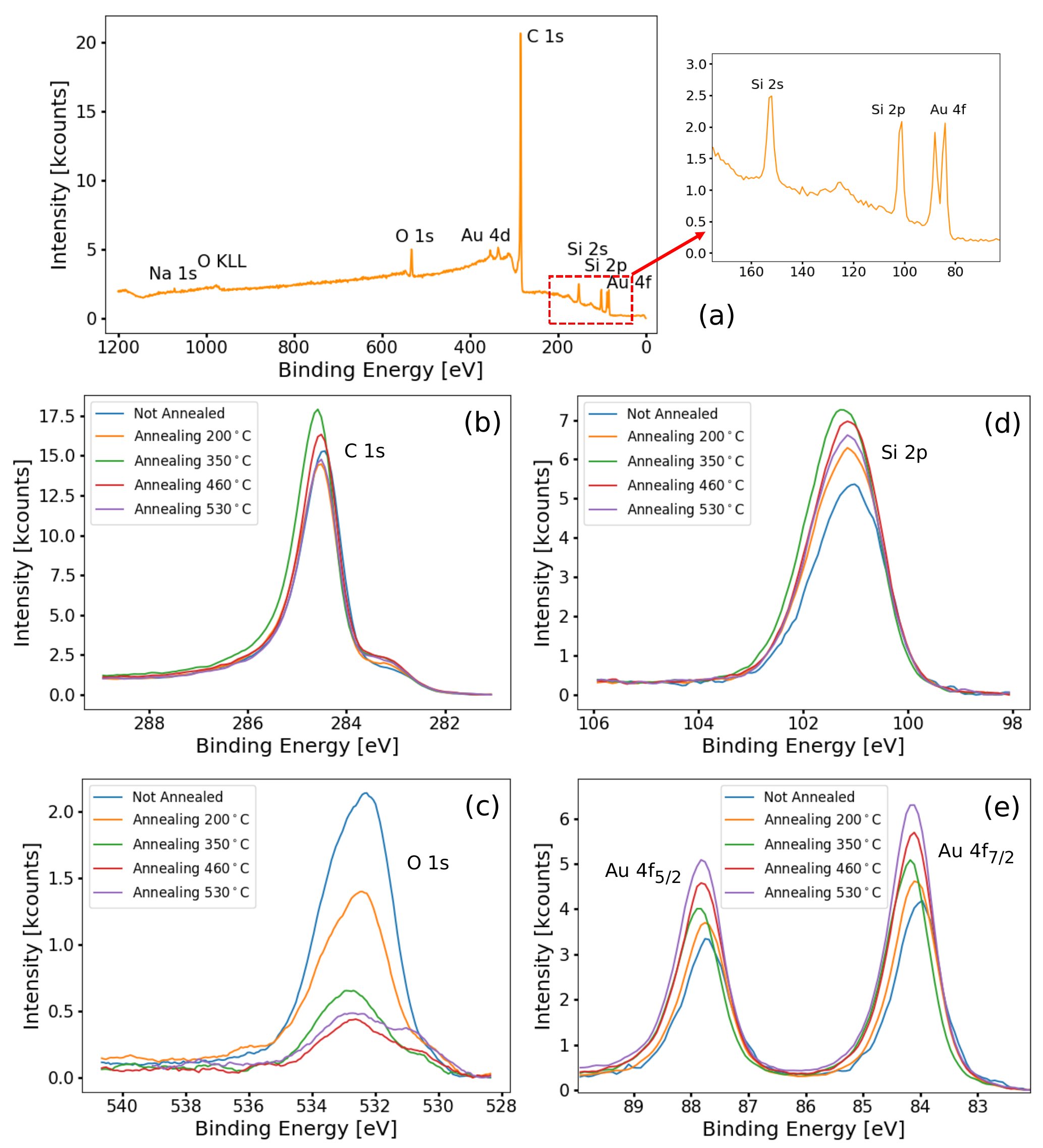}
}
\caption{\label{fig:auxps} (a) Wide scan photoemission spectrum of a AuNP-functionalized 3DG sample prior to annealing. A close up on the Si peaks and the Au 4f peak is also shown. (b), (c), (d), (e) Comparisons of spectra of C 1s, O 1s, Si 2p, and Au 4f, respectively, acquired after annealing the Au-functionalized 3DG sample at different temperatures.}
\end{figure*}

Furthermore, the AuNP-functionalized 3DG has been analyzed by XPS measurements. Figure \ref{fig:auxps}a shows a wide scan photoemission spectrum of a 3DG sample just after functionalization. The C 1s peak is the most prominent, since carbon is the main constituent of the sample surface. Silicon peaks, originating from the SiC underneath, are clearly visible, as well as gold peaks, which confirm the presence of Au on the surface after functionalization. Oxygen is also observed (O 1s peak and the O KLL Auger peak), since the sample has been exposed to air before the XPS measurements. No contaminants are detected, except for a weak Na signal (coming from the sodium citrate cap layer of the AuNPs). After this measurement, a series of annealing steps at increasing temperatures has been performed, followed each time by a new set of XPS measurements. The annealing temperatures were 200, 350, 460, and 530~$^\circ$C.

Figure \ref{fig:auxps}b shows high resolution spectra of the C 1s core level acquired after each of the annealing steps. The spectra remain stable with increasing annealing temperature. 
Spectra are fitted with the three standard components of epitaxial graphene grown on SiC (see Fig.~S3a). The main component, positioned at 284.5~eV, is due to the graphene sp$^2$ C atoms, the left shoulder (at 285.5~eV) due to the buffer layer sp$^3$ C atoms, and the right shoulder (at 283.1~eV) due to C-Si bonds \cite{Trabelsi2017}. 

Figure \ref{fig:auxps}c shows O 1s spectra acquired after each of the annealing steps. Increasing the annealing temperature, the signal sharply decreases. This is due to water desorption and to the degradation of the NPs cap layer. These spectra are fitted with two components (see Fig.~S3b). The higher energy component, centered at 533.2~eV, is assigned to the superposition of \ch{O-C=O} bonds (due to the citrate in the NPs cap layer) and SiO$_2$, while the other, centered around 531.9~eV, to \ch{-OH} (due to adsorbed water) \cite{Sivaranjini2018}. At 350~$^\circ$C of annealing temperature, only one peak, positioned at 533.2~eV, is observed and attributed to SiO$_2$ since cap layer is already decomposed. In addition, above 350~$^\circ$C of annealing temperature, a small signal emerges at 530.8~eV. This is attributed to indium oxide (In is used to attach samples to the sample-holder in the XPS chamber). Indeed, the In 3d peak is also visible in the wide scans after the annealing above 350~$^\circ$C, not reported here.


The Si 2p spectra (cf. Fig.~\ref{fig:auxps}d) are almost unchanged by the annealing. This indicates that no gold silicide forms during the thermal treatments. The signal due to SiO$_2$ is not visible here because, considering the photoionization cross sections of O 1s and Si 2p, and the intensity of the SiO$_2$ component in the O 1s peak, the contribution of SiO$_2$ results to be negligible in the Si 2p spectrum.

Finally, the Au 4f peaks shown in Fig.~\ref{fig:auxps}e increase their intensity with increasing annealing temperature. This is a consequence of the gradual decomposition of the NPs cap layer (which, when present, attenuates the gold signal). Except for the intensity increase, the peaks do not change with temperature. The peak positions are constant with Au 4f$_{7/2}$ at 84.05~$\pm$~0.05~eV, and all spectra are well fitted with fixed branching ratio between the 5/2 and 7/2 components (equal to 3/4). This is a further confirmation of the absence of Si-Au bonds that should arise in the Au 4f photoemission signal shifted to higher energy \cite{Ferrah2022}. In conclusion, AuNPs-functionalized 3DG is thermally stable up to a temperature of at least 530~$^\circ$C.

\subsection{\label{pd1}Functionalization with Palladium Nanoparticles}

\begin{figure*}
\subfloat{\includegraphics[width=0.95\textwidth]{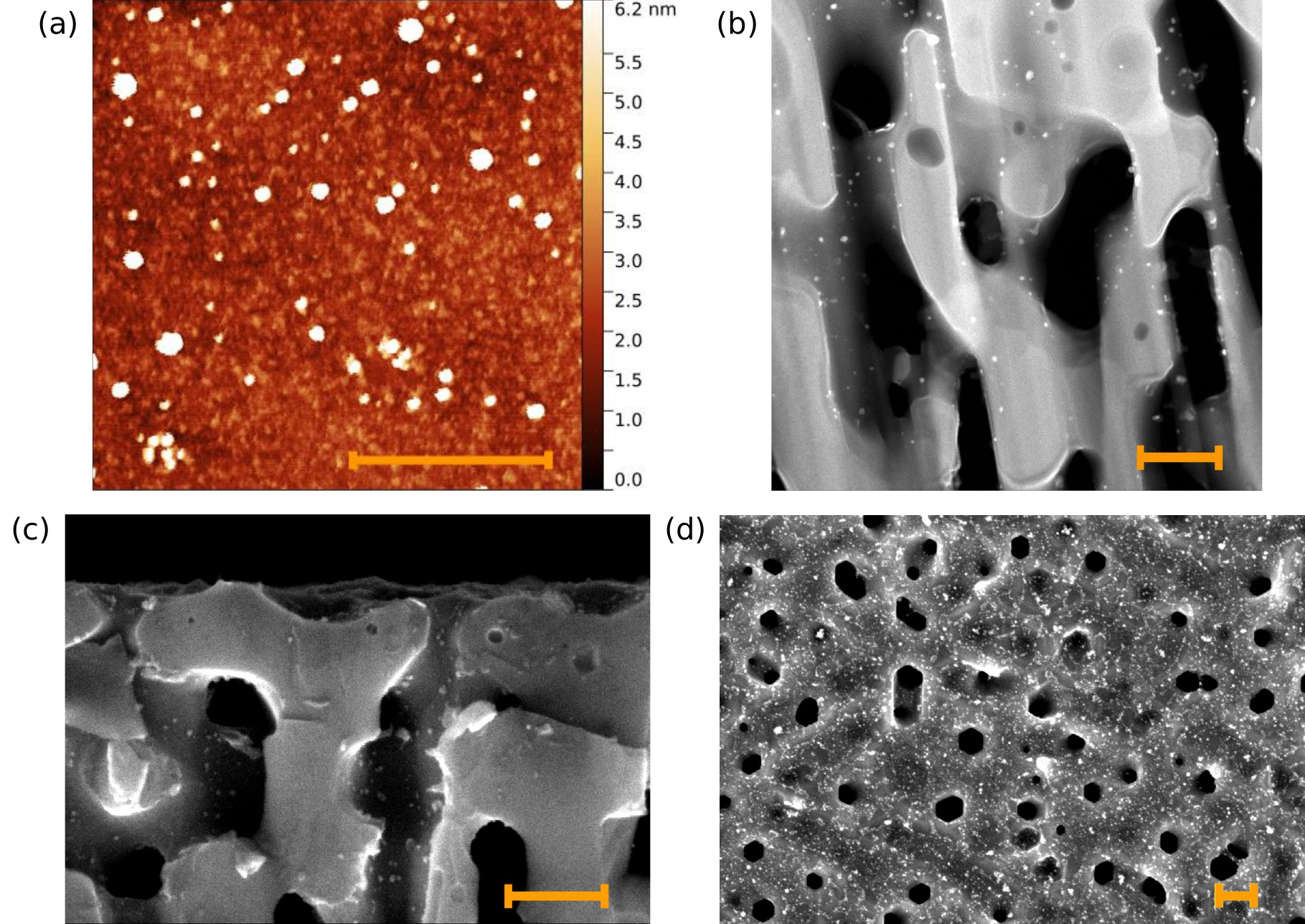}
}
\caption{\label{fig:pvp} (a) AFM image of PVP-PdNPs deposited on a SiO$_2$ substrate. The measurement has been conducted after annealing the sample at 230~$^\circ$C for 4 hours to decompose the NP cap layer. (b) HAADF-STEM cross-sectional image of a 3DG sample functionalized with PVP-PdNPs. PVP-PdNPs appear as bright spots on the walls of the pores. (c), (d) SEM images of a 3DG sample functionalized with PVP-PdNPs in top-view and cross-section, respectively. The scale bar in (a)--(d) indicates 200~nm.}
\end{figure*}

We proceeded to investigate the functionalization with palladium nanoparticles, PdNPs. Palladium was chosen since it is a very versatile metal, with applications in the fields of sensors, as well as energy storage and conversion.

PdNPs were synthesized in situ, following two different procedures described in the Methods Section. The first PdNPs we synthesized and studied were SDS-PdNPs, where SDS refers to sodium dodecyl sulphate involved in the NPs synthesis. After the synthesis, the NPs were dispersed on a SiO$_2$ substrate and characterized by AFM (see Fig.~S4a). The NPs resulted to be spherical and monodispersed, with an average diameter of (8.0$\pm$1.5)~nm.

As for the AuNPs, we proceeded to functionalize a 3DG sample with SDS-PdNPs by immersion in the nanoparticle solution for 24~h. Then, the functionalized sample was analyzed by SEM. The functionalization was successful, since a large number of NPs were detected on the sample. We also observed a non--homogeneous NPs distribution over the 3DG surface, and a tendency to clustering (SEM images of SDS-PdNP-functionalized 3DG are shown in Fig.~S4b). Another aspect which has emerged from Energy-Dispersive X-ray spectroscopy (EDX) measurements performed on this sample is that, at the position of the palladium nanoparticles, in addition to palladium signal, also sulphur was detected. It was still detected after annealing the sample at 800~$^\circ$C (for more details, refer to Fig.~S5). The presence of sulfur is due to the decomposition of the NPs cap layer. Since sulfur poisoning of palladium is well-known to inhibit the palladium catalytic activity, or its capability to adsorb other elements \cite{Gabitto2019, Escandon2008}, this condition should be avoided. Additionally, we observed that the NPs cap layer decomposition produces amorphous carbon regions on the sample surface.

To avoid the SDS-PdNPs clustering and the sulfur poisoning, a new synthesis procedure was adopted (as described in the Methods Section). The new NPs, PVP-PdNPs (where PVP refers to the poly(N-vinyl-2-pyrrolidone) involved in the NPs synthesis), are not subjected to sulfur poisoning, since no sulfur is involved in their synthesis. In addition, PVP-PdNPs possess a cap layer made of shorter molecules compared to SDS-PdNPs, i.e., with less carbon atoms. This implies a reduction in the carbon residues over the sample surface upon thermal degradation of the capping agent.

After the synthesis, PVP-PdNPs were characterized by AFM to determine their dimension, shape, and dispersion. For the measurements, the nanoparticles have been drop-casted on a SiO$_2$ substrate. Figure \ref{fig:pvp}a shows an AFM image of PVP-PdNPs over the SiO$_2$ support. They appear spherical with an average diameter of $(7.2\pm3.0)$~nm. Moreover, NPs clustering is almost absent.

Also with PVP-PdNPs, the functionalization of 3DG was performed by immersing samples in the NPs solution (0.1~mg/ml) for 24 hours. After drying samples in air, HAADF-STEM and SEM measurements were performed. 

Figure \ref{fig:pvp}b shows a HAADF-STEM image acquired in cross-section on a thin lamella of the sample prepared via Focused Ion Beam (FIB) milling. Palladium nanoparticles are clearly visible as bright spots in the HAADF-STEM image. EDX maps (Fig.~S6) confirmed that these spots are composed of Pd. PVP-PdNPs are evenly distributed over the walls of the pores, however, their density decreases with increasing depth, as observed in AuNP-functionalized samples. In the first 750~nm below the surface, we have identified 155 NPs over a 750$\times$750~nm$^2$ scanned area, which corresponds to a density $\rho \simeq 276\ \text{NPs}/\mu\text{m}^2$. While between 0.75~$\mu$m and 1.5~$\mu$m below the surface, we measured a density $\rho \simeq 98\ \text{NPs}/\mu\text{m}^2$. 

Figures \ref{fig:pvp}c and \ref{fig:pvp}d show a SEM image of the top-surface and in cross-section, respectively. Unlike SDS-PdNPs, PVP-PdNPs appear well distributed over the sample surface, and clustering is absent, as shown in Fig. \ref{fig:pvp}c. Even if PVP-PdNPs are smaller than AuNPs, they exhibit a good contrast in SEM images. This was different for SDS-NPs, because their longer capping molecules shielded the Pd, sharply reducing the contrast. This improved contrast allowed to observe PdNPs inside the pores also via SEM, as shown in Figure \ref{fig:pvp}d, and to estimate their density. The nanoparticle top-surface density is $\sigma=(3500\pm740)\ \text{NPs}/\mu\text{m}^2$, while, in the first 1~$\mu$m below the surface, the density inside the pores is $\rho=(172\pm64)\ \text{NPs}/\mu\text{m}^2$. The larger relative errors in the NP densities as compared to the Au NPs are a consequence of the smaller size of the Pd NPs. The NP density inside the pores measured via SEM (computed over a large set of different measurements) is in good agreement with results obtained by HAADF-STEM. This comparison demonstrate the goodness of SEM evaluation, performed in a depth range which comprehends both areas investigated by HAADF-STEM.

Importantly, both HAADF-STEM and SEM measurements demonstrate that PdNP densities are more than a order of magnitude larger than the AuNP densities. This is evidence that smaller nanoparticles are more likely to diffuse inside the pores of 3DG and to stick to their surface.

\begin{figure*}
\subfloat{
\includegraphics[width=0.95\textwidth]{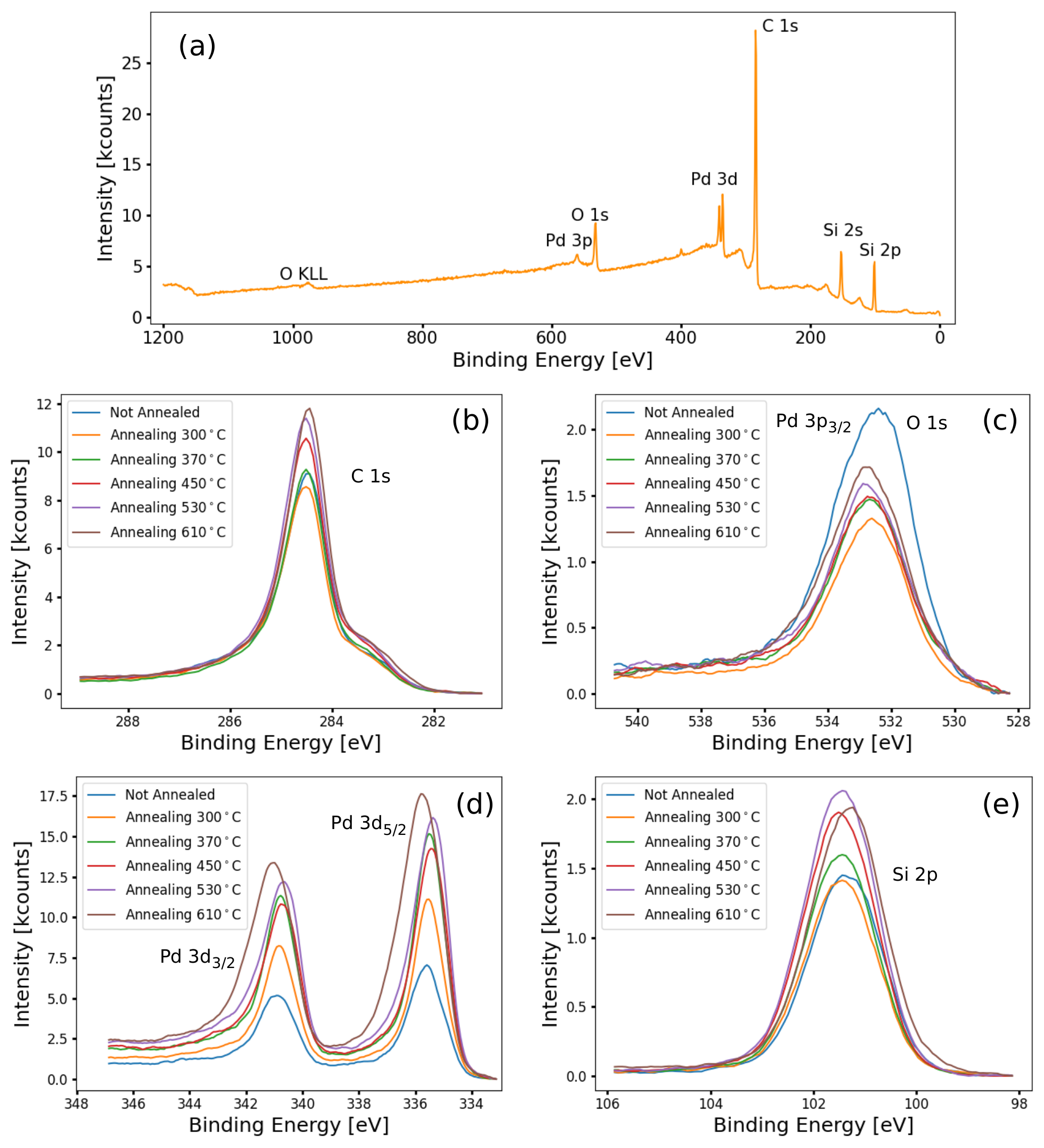}
}
\caption{\label{fig:pdxps} (a) Wide scan photoemission spectrum of a PVP-PdNP-functionalized 3DG sample prior to annealing. (b), (c), (d), (e) Comparison of spectra of C 1s, O 1s and Pd 3p$_{3/2}$, Pd 3d, and Si 2p, respectively, acquired after annealing the Pd-functionalized 3DG sample at different temperatures.}
\end{figure*}

PVP-PdNPs-functionalized 3DG was also analyzed by XPS. Figure \ref{fig:pdxps}a shows a wide scan photoemission spectrum of a 3DG sample just after functionalization. The C 1s peak is the most intense, as for the Au-functionalized sample. Si 2p and 2s peaks are observed, as well. Palladium signal, in particular 3d peaks, exhibits an intensity similar to silicon peaks. Even considering the difference in photoionization cross--section between Pd and Si, this is a clear indication of the large amount of Pd deposited on the sample, confirming the successful functionalization. Also oxygen is detected (O 1s and O KLL Auger peak), consistent with the exposure of the sample to air prior to spectroscopic measurements. Contaminants are not detected. As for the AuNPs-functionalized sample, after the initial set of measurements, a series of annealing steps were performed on the sample. Each annealing step was followed by a new set of XPS measurements. The explored annealing temperatures were 300, 370, 450, 530, and 610~$^\circ$C. 
 
Figure \ref{fig:pdxps}b shows C 1s spectra obtained after each annealing. As for the Au-functionalized sample, variations in the C 1s spectra with increasing annealing temperature are not observed. Information on the fitted spectra of C 1s and following core levels are reported in Fig.~S7.

Figure \ref{fig:pdxps}c shows the photoemission spectra in the binding energy range of both the O 1s and Pd 3p$_{3/2}$ peaks. Therefore, the signal has been fitted with four components, two for oxygen (as for AuNPs-functionalized samples), one for Pd, and one for PdO. 
Prior to annealing, most of the signal is due to oxygen peaks. After the first annealing, a sharp drop in the oxygen signal is observed, compatibly with water desorption (the inset of Fig.~S7b shows the intensity of the components as a function of temperature). Increasing the annealing temperature, the palladium peak, positioned at 532.4~eV, became dominant. Indeed, the PVP cap layer start to decompose around 350~$^\circ$C \cite{Du2006} which results in a reduction in the palladium signal attenuation, and thus in an increase in the intensity of the Pd 3p$_{3/2}$ peak. During the cap layer decomposition, part of the released oxygen atoms are available for Pd oxidation. Therefore, the PdO contribution is negligible up to annealing temperature of 370~$^\circ$C.

Figure \ref{fig:pdxps}d shows Pd 3d spectra at different annealing temperatures. Similarly to the Pd 3p$_{3/2}$ peak, the Pd 3d peak intensity increases with annealing temperature. In addition, after annealing at 610~$^\circ$C, the Pd 3d core level emission shifts by 0.2~eV to higher energy. This is understood by fitting the core level with two components for palladium metal and palladium silicide \cite{Krause2019} (see Fig.~S7c). The palladium metal 3d$_{5/2}$ peak is centered at 335.5~eV, while for the silicide it is 1.1~eV higher. Palladium silicide signal is negligible up to 530~$^\circ$C. Instead, after the last annealing step at 610~$^\circ$C, it became non--negligible, resulting in a blue-shift of the peaks.

The silicide formation is further confirmed by the Si 2p spectra, which are shown in Figure \ref{fig:pdxps}e. Similarly to Pd 3p, Si 2p peak position is constant at 101.2~eV with increasing annealing temperature up to 530~$^\circ$C, while, after the last annealing step, the peak is red-shifted by 0.2~eV. This is compatible with the appearance of a lower energy peak, centered at 100.6~eV, due to Pd$_2$Si formation \cite{Dai1995}.

In conclusion, PVP-PdNP functionalization of 3DG is a clean process (no contaminants are detected) which leads to a high coverage of NPs over the sample surface. In addition, the as-produced samples are observed to be thermally stable up to a temperature of 530~$^\circ$C, above which silicide formation indicates a reaction of the PdNPs with the substrate.

\section{\label{conclusion} Conclusions}

We have demonstrated a method for the successful functionalization of epitaxial 3DG with metal (gold and palladium) nanoparticles. This method does not involve any chemical reaction at the substrate surface, and the functionalization is driven by the nanoparticle diffusion and their subsequent sticking to the pore walls via electrostatic and van der Waals interactions. Therefore, our approach can be applied to any kind of nanoparticles, even non-metallic, or in shapes different from spherical. The only limitation is the diffusivity of different kinds of nanoparticles inside the porous region, but, as we demonstrated, the diffusion can be promoted by utilizing smaller NPs or larger pores. 
Considering the ease in obtaining a successful functionalization and the wide range of usable nanoparticles, our results pave the way to numerous future studies on this new class of materials. 

We demonstrated the fundamental role of graphene in the functionalization. In fact, if a sample exhibits a low graphene coverage, or a low quality of the 3DG, a sharp reduction in NP density was observed.

Through XPS measurements we established that the method is clean, and no contaminants are detected over the functionalized 3DG samples. XPS provided information on the thermal stability of the system. 3DG functionalized with gold or palladium is stable up to at least 530~$^\circ$C. This feature could be exploited for high temperature catalytic reactions.

Through EDX we were able to detect the sulfur poisoning of palladium in SDS-PdNPs. We observed that the poisoning was not removed by annealing the sample at 800~$^\circ$C. Therefore, we adopted a different synthesis routine which does not involve sulfur. The NPs synthesized following this new recipe also guaranteed much less clustering, a more homogeneous distribution over the 3DG surface and, because of their shorter capping molecules, a reduction in the residual carbon from the thermal degradation of the capping molecules.

In conclusion, we realized a very versatile support which could be employed in a wide range of fields. Au-decorated 3DG could be used as substrate for surface-enhanced Raman spectroscopy. Pd-functionalized 3DG could achieve great performance in hydrogen sensing and storage. In general, this new class of materials is suitable for a multitude of catalytic applications, even in harsh environments, which can be tuned by selecting an appropriate metal for the specific reaction. For example, using platinum NPs for the functionalization could be useful for fuel cell electrodes. Besides, it has to be noticed that these materials could also have applications in biology since graphene and SiC are bio-compatible.

\section{\label{exp}Methods}

\subsection{\label{sec:psic}Porous Silicon Carbide}

In the present study n-doped 4H-SiC wafers are used (purchased from SiCrystal). They are 350~$\mu$m thick and possess a bulk resistivity of 0.02~$\Omega$cm. All porosification processes are performed on the (0001) Si-face of the SiC wafer.

The porosification of the wafers is achieved by photoelectrochemical etching. The etching procedure is composed of three steps; first Metal-Assisted PhotoChemical Etching (MAPCE) \cite{Leitgeb2017}, then Photo ElectroChemical Etching (PECE), and finally, a second MAPCE step.
Prior to the first MAPCE step, wafers are cleaned by inverse sputter etching in a LS730S Von Ardenne sputter equipment for 150~s. Without breaking the vacuum, 300~nm thick platinum pads are sputter deposited at the wafer edges (these act as local cathodes during the etching). Then, the sample is transferred to an electrochemical cell for a 30 minutes MAPCE. Here, the utilized etching solution contains 1.31~mol/l HF and 0.15~mol/l H$_2$O$_2$. The cell is equipped with a 18~W UV source. The interface between the SiC and the solution acts as cathode. Here, SiC oxidises to SiO$_2$ which is dissolved by HF, resulting in the formation of pores over the sample surface. 
The PECE solution is composed by 1200~ml of deionized water, 150~ml of HF (48 \%) and 150~ml of pure ethanol. The same UV lamp as in MAPE is utilized, and a bias of 11.5 V is applied between the two sides of the sample. Because of the applied bias, PECE produces a more directional and deeper porosification.
To guarantee a homogeneous porosification \cite{Leitgeb2017a}, after PECE, a second MAPCE is performed in the same conditions as the first step, except for a longer time of 1 hour.

\subsection{\label{sec:3dg growth}Three-dimensional Graphene Growth and Characterization}

The porousified SiC wafers are cut into pieces (2$\times$8~mm$^2$) by a wafer saw. Pieces are individually transferred to an ultra-high vacuum (UHV) chamber where they are degassed overnight at 1000~K. After the degassing, samples are cooled to room temperature (RT) and heated again to 1650~K to allow the graphene formation on the surface of the pores. After 150~s of annealing, the sample is let to cool down to RT.

Veronesi et al. observed that the temperature reached by the sample during the annealing depends on the quality of the mechanical contact between sample and sample-holder \cite{Veronesi2022}. Therefore, by clamping one side of the sample more firmly than the other, we obtained the non--homogeneous graphenized sample discussed in Section \ref{sec:aunps}.

The 3DG was characterized by Raman spectroscopy, which allowed to demonstrate the presence of monolayer graphene, and to determine its strain state and doping level (refer to Fig.~S1). Scanning Tunneling Microscopy was also performed. Atomically resolved images further confirm the presence of graphene on the top surface of the porous material (for more details, refer to Fig.~S8). To characterize the sample morphology, both High-Angle Annular Dark-Field Scanning Transmission Electron Microscopy (HAADF-STEM) and Scanning Electron Microscopy (SEM) were utilized. From HAADF-STEM measurements we observed an average pore transverse dimension of (149$\pm$48)~nm. Through SEM, both the sample surface and the cross-section (after cleaving the sample with a diamond tip) were investigated. From SEM measurements we determined an average pore transverse dimension of (166$\pm$20)~nm, consistent with the HAADF-STEM measurement, and an average distance between pores of (180$\pm$29)~nm. Due to the large set of SEM measurements taken into account for extracting these values, their relative errors are lower compared to those from HAADF-STEM. The average depth of the porous region, extracted from SEM measurements performed on different samples, is (18.1$\pm$0.3)~$\mu$m (for SEM images refer to Fig.~S9).

\subsection{\label{sec:sds}Palladium Nanoparticle Synthesis}

All chemicals were purchased from Sigma-Aldrich unless otherwise indicated, and were used without further purification. Ultrapure water purified using a Millipore Milli-Q water system was used to prepare aqueous solutions.

SDS-PdNPs are synthesized from an aqueous solution of palladium(II) acetate (Pd(OAc)$_2$, 98\% pure), and sodium dodecyl sulphate (SDS, $>$99\% pure).
SDS is dispersed in MilliQ water with concentration 0.05~M. Here, 10~mg of Pd(OAc)$_2$ are dissolved. The obtained solution is transferred to a three-neck flask equipped with a reflux condenser. The solution is heated to 100~$^\circ$C under magnetic stirring for 6 hours. The heat induces the thermal decomposition of SDS to 1-dodecanol. This, in turn, induces the formation of PdNPs upon reduction of Pd$^{2+}$ ions, and 1-dodecanol oxidation to dodecanoic acid which acts as stabilizer for the produced nanoparticles \cite{Karousis2008}. 
Once the reaction is terminated, the solution is cooled to RT. PdNPs are collected by ultrahigh speed centrifugation (13000 rpm) for 10 minutes and washed sequentially with water and ethanol (supplied by Carlo Erba Reagents) to remove the excess of surfactant. Finally, SDS-PdNPs are dispersed in ethanol and stored at $-20$~$^\circ$C.

PVP-PdNPs are synthesized following the polyol method in which an alcohol is used for the reduction of the metal precursor \cite{Favier2019}. A capping agent is required to prevent the clustering of the nanoparticles. Poly(N-vinyl-2-pyrrolidone) (PVP) has been chosen as capping agent for our synthesis since it is known that that the carbonyl groups of PVP partly coordinate to the surface Pd atoms of the Pd nanoparticles \cite{Nemamcha2005}. Furthermore, part of the main chain of PVP is expected to be adsorbed on the surface Pd atoms by hydrophobic interaction.

In this synthesis, 0.0205~g (0.09~mmol, 10~mM) of Pd(OAc)$_2$ are dissolved in 1.5~ml of ethylene glycol (EG, $>$99\% pure) under magnetic stirring for 2 h, in ambient conditions. During the process, the solution turns from light orange to dark brown, indicating the formation of PdNPs. In a second vial, 0.05~g of PVP (average molecular weight 40,000) are dissolved in 3~ml of EG and stirred for 2~h.
In a three-neck flask, equipped with a reflux condenser, 5~ml of EG are poured. The EG in the flask is heated to 160~$^\circ$C under magnetic stirring. Aliquotes of the PdNPs solution (30~$\mu$l) and the PVP solution (60~$\mu$l) are injected into the flask every 30~s (50 additions in total). Then, the flask is cooled to RT and the obtained suspension is centrifuged at 13000~rpm for 10 minutes to remove the precipitate containing large nanoparticles (diameter above $\sim$ 80~nm). The smaller nanoparticles (diameter $\lesssim$ 10~nm) are instead suspended in the supernatant. These are collected by dilution of the solution with acetone ($>$99\% pure) and centrifuging. The product is washed with acetone three times to remove most of the EG and the excess of PVP. Finally, PVP-PdNPs are dispersed in ethanol and stored at $-20$~$^\circ$C.

\subsection{\label{sec:measurements}Measurements}

Raman spectroscopy and STM characterization of pristine 3DG samples were performed utilizing an inVia confocal Raman microscope (equipped with a 532~nm laser, with spot diameter of 1~$\mu$m) from Renishaw, and a VT-UHV STM from RHK, respectively. 

The AFM characterization of PdNPs (over SiO$_2$ substrates) was conducted in tapping mode with a Dimension Icon from Bruker.

Functionalized sample has been investigated in cross-section via High Angle Annular Dark Field - Scanning Transmission Electron Microscopy. The analysis has been conducted using a X-Ant-EM Titan microscope operating at 300 kV, equipped with an aberration corrector and a monochromator. Prior to HAADF-STEM measurements, the sample has been milled into lamellas via Focused Ion Beam.

For SEM measurements conducted on functionalized 3DG sample, a FEG-SEM Merlin from ZEISS was used.

The EDX measurements of sulfur poisoning of palladium were performed with a QUANTAX EDS from Bruker (mounted to an Ultra Plus FE-SEM from ZEISS). 

XPS characterization of 3DG samples functionalized with AuNPs and PVP-PdNPs was performed using a Surface Science Instrument SSX-100-301 spectrometer operating an Al K$\alpha$ x-ray source with overall energy resolution of 0.9~eV.

\bibliography{bib}

\end{document}